\documentclass[aps,reprint,prl]{revtex4-1}
\usepackage[utf8]{inputenc}
\usepackage{mathtools}

\usepackage{graphicx}
\usepackage{dcolumn}
\usepackage{bm}
\usepackage{xcolor}
\usepackage{amsmath}
\usepackage{amssymb}
\usepackage{hyperref}

\begin{document}
\title{Descendant distributions for the impact of mutant contagion on networks}


\date{\today}
\author{Jonas S. Juul}
\email{jonas.juul@nbi.ku.dk}
\affiliation{\mbox{Niels Bohr Institute, University of Copenhagen, Blegdamsvej $17$, Copenhagen 2100-DK, Denmark}}
\affiliation{Center for Applied Mathematics, Cornell University, Ithaca, New York 14853, USA}
\author{Steven H. Strogatz}
\email{strogatz@cornell.edu}
\affiliation{Center for Applied Mathematics, Cornell University, Ithaca, New York 14853, USA}

\newcommand{\ER}{Erd\H{o}s--R\' enyi{ }}

\begin{abstract}
Contagion, broadly construed, refers to anything that can spread infectiously from peer to peer.  Examples include communicable diseases
, rumors, misinformation, ideas, innovations, bank failures, and electrical blackouts. Sometimes, as in the 1918 Spanish flu epidemic, a contagion mutates at some point as it spreads through a network. Here, using a simple susceptible-infected (SI) model of contagion, we explore the downstream impact of a single mutation event. Assuming that this mutation occurs at a random node in the contact network, we calculate the distribution of the number of  ``descendants,'' $d$, downstream from the initial ``Patient Zero" mutant. We find that the tail of the distribution decays as $d^{-2}$ for complete graphs, random graphs, small-world networks, networks with block-like structure, and other infinite-dimensional networks. This prediction agrees with the observed statistics of memes propagating and mutating on Facebook, and is expected to hold for other effectively infinite-dimensional networks, such as the global human contact network. In a wider context, our approach suggests a possible starting point for a mesoscopic theory of contagion. Such a theory would focus on the paths traced by a spreading contagion, thereby furnishing an intermediate level of description between that of individual nodes and the total infected population. We anticipate that contagion pathways will hold valuable lessons, given their role as the conduits through which single mutations, innovations, or failures can sweep through a network as a whole.
\end{abstract}
\maketitle

\section{Introduction}
The concept of contagion began in epidemiology, where it was used to describe the spread of disease between people in close contact. Nowadays contagion has taken on a broader meaning; it refers to any sort of process that can spread infectiously from node to node through a network~\cite{newman2018networks,watts2002simple,dodds2004universal, castellano2009statistical,christakis2007spread,sayama2015social,lehmann2018complex}.  Along with  communicable diseases~\cite{pastor2001epidemic,anderson1991infectious, keeling2011modeling,de2016physics,fennell2019multistate,pastor2015epidemic,castellano2010thresholds,kiss2017mathematics}, examples of contagions include rumors~\cite{daley1965stochastic}, misinformation~\cite{Vosoughi1146},  ideas~\cite{bettencourt2006power}, innovations~\cite{mellor2015influence,juul2019hipsters,krapivsky2011reinforcement}, bank failures~\cite{ haldane2011systemic}, and electrical blackouts~\cite{brummitt2012suppressing}.

When a contagion spreads, 
it propagates from one or more ``parent'' nodes to a  number of ``descendant'' nodes. 
Enumerating the descendants in all the paths stemming from a parent can reveal important and useful information. 
In particular, suppose a contagion mutates into a more pernicious form as it travels. Then counting its descendants would tell us how many  nodes will be confronted by this nastier strain. A mutation event of this sort occurred in $1918$, and gave rise to the Spanish flu epidemic that killed millions of people worldwide~\cite{andreasen2008epidemiologic}. Similar (but less consequential) mutations happen online when users modify memes to make them funnier or stickier before sharing them with their peers~\cite{adamic2016information}.

Here, we derive exact results for the impact of a single mutation event, assuming the contagion dynamics are governed by the so-called susceptible-infected (SI) model. Our goals are to understand, in a statistical sense, how many nodes will ultimately get infected by the mutant strain, and to clarify how the results depend on the structure of the underlying contact network. 


\section{Descendant distributions}
To make analytical progress, we consider an extremely simplified model in which each node is either susceptible or permanently infected (Fig.~1). This SI model effectively assumes infinite transmissibility of the contagion, and ignores the possibility of recovery, death, migration, vaccination, temporary immunity, latency periods, heterogeneity of susceptibility and infectiousness, and many other realistic considerations. All of these would make for interesting extensions of our work.

As the contagion spreads (Fig.~1(a)), we record which nodes caught it from which, and plot the resulting paths of infection as an epidemic tree (Fig.~1(b)). Then we count how many nodes would be affected by a mutation occurring at a random ``Patient Zero'' node. In the example shown in Fig.~1(c), the mutant infection occurs at node $B$ and is passed along to the two nodes below it. Of course, if the mutation had occurred elsewhere, it could have produced either more descendants (e.g., three descendants, had the mutation occurred at $A$) or fewer (zero descendants, had it occurred at $C$). Thus, the natural statistical quantity to study is the \emph{distribution} of the number of descendants, aggregated over all possible Patient Zero nodes.  

\begin{figure*}
    \centering
    \includegraphics[]{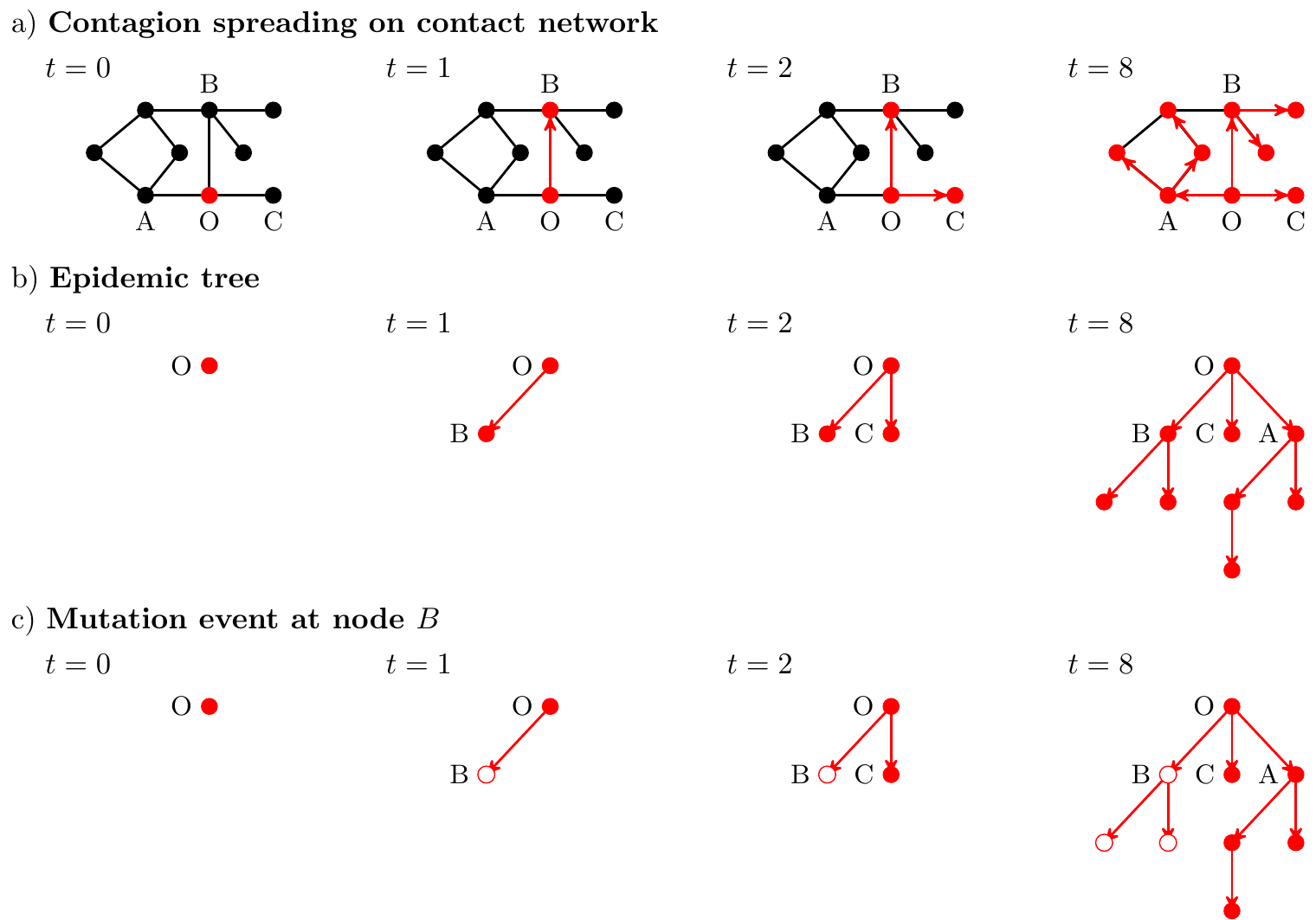}
    \caption{\textbf{Simple SI model of contagion spreading on a network and its corresponding epidemic tree.} Black filled circles denote susceptible nodes; red filled circles, infected nodes; red open circles, nodes infected by a mutant strain of the infection.  \textbf{a)} Starting with a single infected seed $O$ at time $t=0$, another node gets infected at random at the next time step. Any edge between an infected node and a susceptible node has an equal chance of  being the next edge over which the contagion spreads. We keep track of which nodes transmitted and received the infection at every time step, until ultimately every node is infected. \textbf{b)} The epidemic tree shows who infected whom in the contagion process depicted in \textbf{a)}. We draw this tree with the seed on top. The nodes that the seed infected are drawn in the second layer, and so on. A descendant of node $i$ is defined as any node that directly or indirectly received the infection from node $i$. Such a descendant node $j$ can be reached by starting at node $i$ and following a sequence of directed edges downward through the epidemic tree until the path ends at $j$. \textbf{c)} If a mutant infection occurs at some node ($B$, in the example shown here), that node passes the mutated strain on to all its descendants (two descendants, in this example). }
    \label{fig:epidemic_tree}
\end{figure*}

In one sense, the dynamics assumed here are trivial: one node after another gets infected until no susceptibles remain. But what is not trivial are the descendant distributions implied by the model, as they also depend on the network's structure. 

One limiting case is already understood. In a completely structureless, well-mixed population, the impact of a single mutation can be quantified by the classical stochastic process known as the Yule process. In that case, the probability that a mutant generates exactly $d$ descendants is
\begin{equation}
    P_d = \frac{1}{(d+2)(d+1)}.
\end{equation}
To the best of our knowledge, however, it has been an open problem to extend this result to \emph{structured} populations. 

To learn what to expect, we first compute descendant distributions numerically from Monte Carlo simulations~\footnote{All scripts necessary to reproduce the simulated results are available at \href{https://sid.erda.dk/wsgi-bin/ls.py?share_id=F8JmKmQryb}{https://sid.erda.dk/wsgi-bin/ls.py?share\_id=F8JmKmQryb}.}. For a given random realization of the SI contagion process on a given network, like the one shown in Fig.~1(b), we count the number of descendants of each node and compile a histogram. This histogram, however, merely gives the descendant distribution for one realization of the stochastic dynamics. To extract a more robust statistical measurement, we average over the random location of the initially infected seed node, as well as the random decisions of whom to infect at each step, to obtain an average descendant distribution. 

Figure 2 shows the average descendant distribution for the simplest possible network structure: a complete graph, in which each node is connected to all the others. The downward slope of the plot indicates that many nodes have few descendants, and a few nodes have many descendants. 
Of course, the seed $O$ has every other node as its descendant, as an artifact of the assumed initial conditions. Its corresponding data point in Fig.~2 lies off the curve for this reason.    
\begin{figure}
    \centering
    \includegraphics[width=89mm]{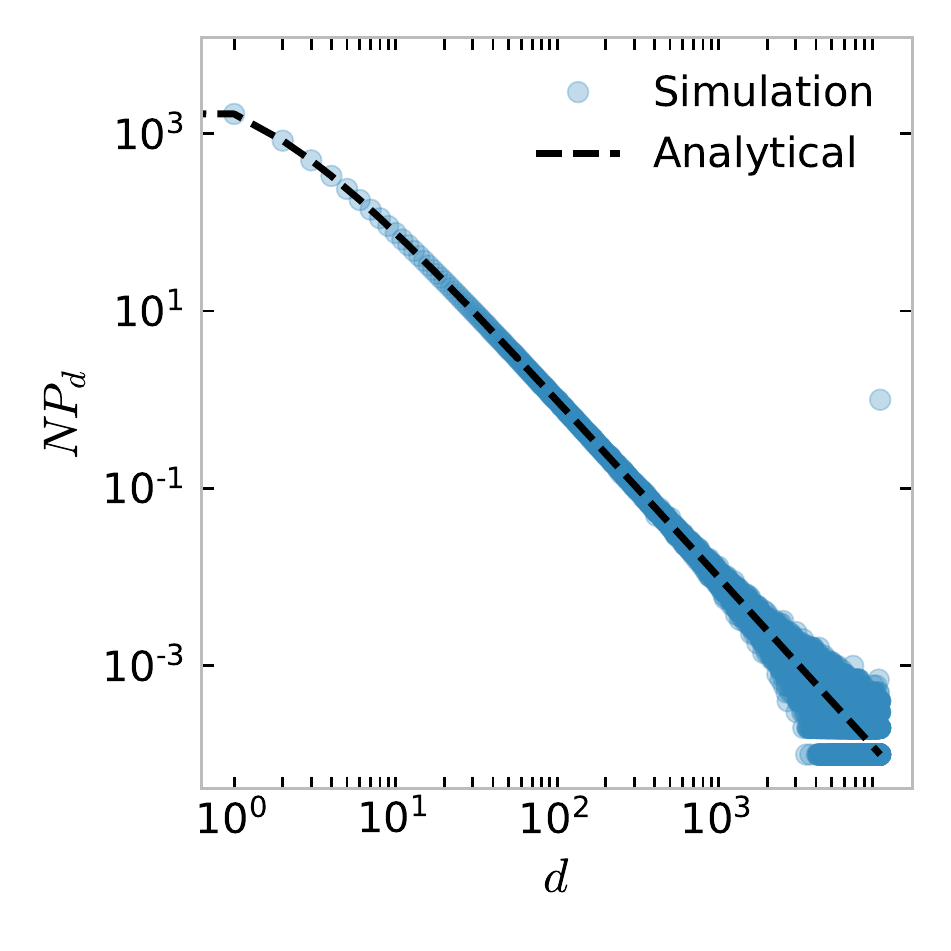}
    \caption{\textbf{Descendant distribution for the SI contagion process on a complete graph.} We simulated the SI model on complete graphs of $N=10^4$ nodes, and averaged the resulting descendant distributions over $10^3$ realizations of the random contagion process, each of which started with a single seed node. Filled circles show the numerically computed distribution of the number, $d$, of descendants of each node in the network. This distribution quantifies the impact that a mutant infection would have on the rest of the population, had it started at a random ``Patient Zero" node. The dashed line shows the analytical result~\eqref{eq:complete_graph_solution}. For large values of $d$, the descendant distribution declines proportional to $d^{-2}$.}
    \label{fig:complete_graph}
\end{figure}

The most striking feature of the descendant distribution in Fig.~2 is its apparent power-law decay for $d \gg 1$. To explain this scaling law intuitively, recall that one way of getting power-law distributions is through  rich-get-richer effects~\cite{yule1925ii,simon1955class,price1976general,barabasi1999emergence}, and observe:
\begin{itemize}
    \item[(i)] If node $i$ infected node $j$, the ancestors of $j$ will be $i$ \emph{and} all the ancestors of $i$.
    \item[(ii)] A node $i$ can acquire a new descendant $j$ if it passes the infection on to $j$, or if one of its descendants passes the infection on to $j$.
\end{itemize}
The first point means that our model contagion process is equivalent to a network that grows by node copying~\cite{krapivsky2005network}. The second point suggests that the probability of a node acquiring more descendants should grow, loosely speaking, in proportion to the number of descendants it already has, thereby making the rich richer.

To sharpen this intuition, we calculate the descendant distribution $P_d$ analytically for some exactly solvable networks~\footnote{See Supplemental Material at [URL will be inserted by publisher] for details of the analytical derivations, additional simulations, and further discussion of the breakdown of the assumption of infinite-dimensionality.}. First, for a complete graph in the limit $N\rightarrow \infty$, we recover the classical result of Yule,
\begin{equation}
P_d
= \frac{1}{(d+2)(d+1)}.
\label{eq:complete_graph_solution}
\end{equation}
This result was also found by Krapivsky and Redner for the in-degree distribution of networks growing by node copying~\cite{krapivsky2005network}. Figure~2 shows that this result agrees well with our simulation data. For further discussion of the connection between the Yule process and rich-get-richer effects, see Ref. \cite{pachon2016random}.
\begin{figure}
    \centering
    \includegraphics[width=89mm]{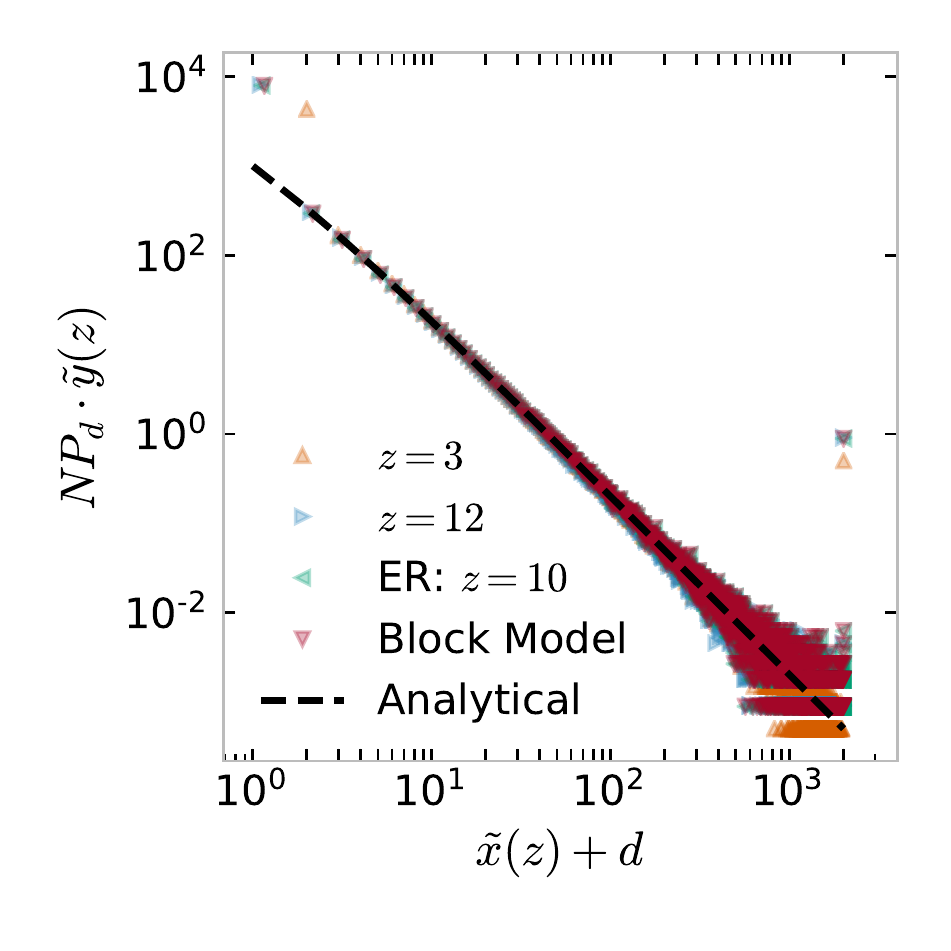}
    \caption{\textbf{Descendant distributions for the SI contagion process on random networks.} We simulated the SI process on $z$-regular configuration models, \ER (ER) networks, and networks with block structure of $N=10^4$ nodes. The block network has 4 equally-sized \ER blocks and parameters $z=8.25$ and $p=0.002$. The descendant distributions have been rescaled to collapse on the analytical solution~\eqref{eq:ER_analytical}. This rescaling involved adding $\tilde{x}(z) = (z-1)/(z-2)$ to $d$, and multiplying $P_d$ by $\tilde{x}(z)^{-1}$, the inverse of the scaling factor of $P_d$.} 
    \label{fig:ER}
\end{figure}

Likewise, for several classes of random networks, the descendant distributions can be derived in the limit of infinite network size. For $z$-regular configuration models and \ER random graphs with average degree $z$, we obtain~\cite{Note2} the infinite-$N$ solution
\begin{equation}
P_d = \frac{z-1}{z-2}B\left( \frac{z-1}{z-2}+d,2 \right),
\label{eq:ER_analytical}
\end{equation}
where $B(a,b)$ denotes the beta function. Importantly, this expression reduces to the complete-graph solution for large values of $z$.

More complicated network structures yield similar results~\cite{Note2}. For networks consisting of \ER ``blocks'' with mean degree $z$, and with probability $p$ of connecting each node to a node chosen uniformly at random from nodes located in other blocks, we obtain the solution
\begin{equation}
    P_d = \frac{z-1+p}{z-2+p}B\left( \frac{z-1+p}{z-2+p}+d,2 \right).
    \label{eq:block_analytical}
\end{equation}

Figure~3 shows the simulation results for $z$-regular configuration models, \ER random graphs, and modular networks with block structure, all of size $N=10^4$.
When plotted in a manner suggested by Eqs.~\eqref{eq:ER_analytical} and \eqref{eq:block_analytical}, the simulation data for the different random networks collapse onto a single curve (Fig.~3), consistent with the analytical approximation. 

Finally, for a small-world network created by inserting random shortcuts in a ring lattice, with probability $p$ of connecting a node with a node chosen uniformly at random~\cite{newman1999renormalization}, the analytical solution~\cite{Note2} is
\begin{align}
P_d = \frac{2p+1}{2p} B\left( \frac{2p+1}{2p}+d,2 \right).
\label{eq:SW_solution}
\end{align}
This result agrees well with  simulations; see Fig. 1 in the Supplemental Material~\cite{Note2}. 
\begin{figure}
    \centering
    \includegraphics[width=89mm]{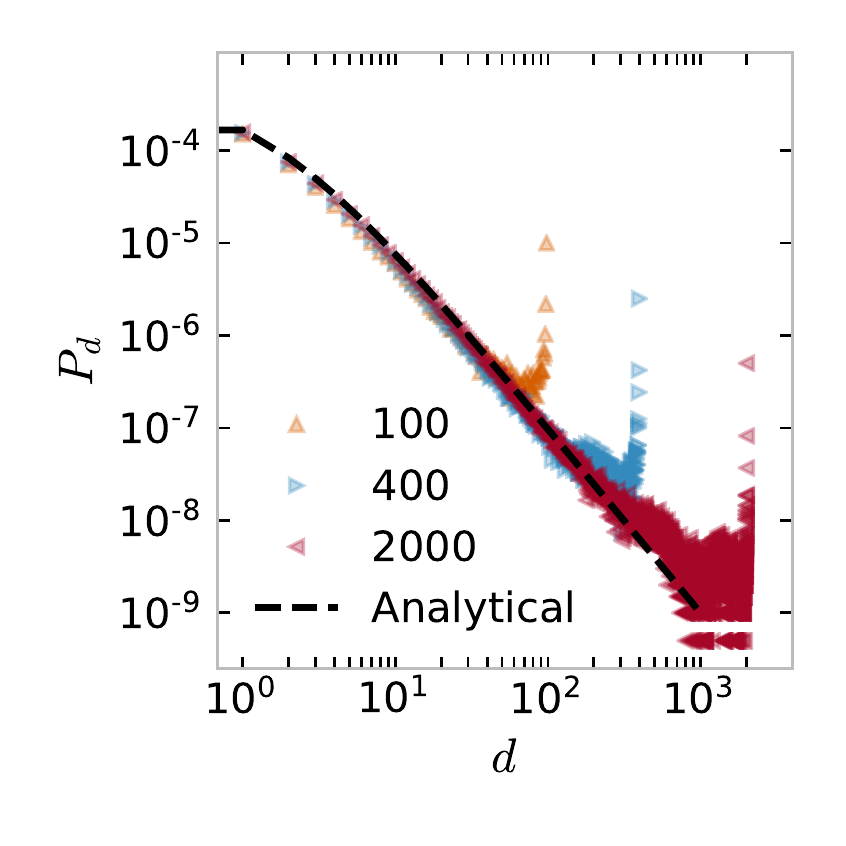}
    \caption{\textbf{Descendant distributions of an SI process simulated on a Facebook subgraph.} We ran simulations of the SI contagion process on a network of merged Facebook ego networks~\cite{leskovec2012learning,snapnets}. Taken together, this  subnetwork  contains  4039 nodes  and  88234  edges. We started a contagion at a random seed node, and ran the simulation until exactly a predefined number of nodes were infected. Then we stopped the spreading, obtained the descendant distribution for the realization, and started a new simulation with a seed chosen uniformly at random. The descendant distributions shown here have predefined cascade sizes $100$, $400$, and $2000$, and are averaged over $10^3$ simulations for each cascade size. The smallness of the subnetwork gives rise to conspicuous finite-size effects in the tail of the distribution. Apart from these effects, the descendant distribution falls on the curve expected for highly connected infinite-dimensional networks, here illustrated with the analytical solution for the descendant distribution for contagion on the complete graph. 
    }
    \label{fig:real_networks}
\end{figure}

\section{Scaling law for the tail}
All the descendant distributions we have calculated so far turn out to decay asymptotically according to the same power law:
\begin{equation}
     P_d  \propto  d^{-2}
\end{equation}
for $d \gg 1$. Further analysis~\cite{Note2} indicates that this inverse-square scaling follows from a property that the complete graph shares with the random networks: they all become infinite dimensional as $N \rightarrow \infty$. (Here, we consider a network to be infinite dimensional if the area and volume of a ball of radius $s$ grow equally fast with $s$; in this context, a ball of radius $s$ is defined as the set of all nodes within $s$ hops of a given node. See Section V in Supplemental Material for details~\cite{Note2} as well as Refs. \cite{nowotny1998dimension,shanker2007graph}) for further discussion of the concept of network dimension.)

On this basis, we expect that the same $d^{-2}$ scaling should hold for other infinite-dimensional networks, but not for one-dimensional chains, two-dimensional grids, three-dimensional lattices, or other networks whose dimensionality remains finite and sufficiently small as the number of nodes tends to infinity. In some sense, this expectation is natural: there are well-known analogies between epidemic models and percolation models, and for many of these, the critical properties vary with dimension for a range of intermediate dimensions and then agree with mean-field theory above some upper critical dimension \cite{ben2000diffusion,pastor2015epidemic,newman2018networks}. Simulations of the model contagion on two-dimensional square grids support this predicted dependence on dimension: descendant distributions deviate significantly from the $d^{-2}$ scaling~\cite{Note2}. Interestingly, scale-free networks also show a departure from the scaling predicted above, but their descendant distributions  merge with our predictions past a crossover, producing the same inverse-square decay in the tail~\cite{Note2}.     

Conveniently, many real-world networks   are effectively infinite dimensional. Consider the social network Facebook, which as of June 2019 had more than 2.4 billion active users. In a fascinating study, Adamic et al.~\cite{adamic2016information} examined memes spreading from friend to friend on the Facebook social graph. Typically, memes would propagate from one user to another without being altered, but occasionally a user would change the content of the meme before resharing it. This would make a new variant of the meme, which would then spread on the network along with previously existing copies. Adamic et al.~\cite{adamic2016information} examined the frequency of different variants of rarely-changing memes, and found that the frequency distribution of the most widely shared variants followed an inverse-square law. Specifically, they found the exponent to be $-2.01\pm0.15$. This exponent matched the prediction of a mean-field model (the Yule process), but it remained unclear why a model without any underlying network structure could account for the exponent obtained from the actual Facebook network. 

Our work suggests that the observed exponent of $-2$ is a consequence of the approximate infinite-dimensionality of the Facebook network. Indeed, Fig.~\ref{fig:real_networks} shows that when we simulate our simple contagion process on a small subnetwork of Facebook~\cite{leskovec2012learning,snapnets}, the resulting descendant distributions match what we would expect for highly-connected infinite-dimensional networks. In particular, apart from effects caused by the small size of the subnetwork, an approximate power-law tail with a slope close to $-2$ emerges.  

\section{Discussion}

The epidemic trees analyzed in this paper, along with their associated pathways of contagion, have been studied previously in diverse disciplines. They have been called adoption paths~\cite{juul2019hipsters}, dissemination trees~\cite{oh2018complex,liben2008tracing}, spreading patterns~\cite{jang2018computational}, causal trees of disease transmission~\cite{vazquez2004causal}, diffusion structure patterns~\cite{zhang2016dynamics}, the structure of diffusion events~\cite{goel2015structural}, and epidemic trees~\cite{haydon2003construction}. 
We have chosen to adopt the term ``epidemic trees,'' although it comes with a significant caveat: Generally the graph of the propagation paths for a contagion need not be a directed tree; in the case of a complex contagion~\cite{monsted2017evidence}, where each child node has two or more parents, the graph could be a directed graph with no cycles. But for the simple contagions studied here, where each child is assumed to have only one parent, the graph of the propagation paths is always a tree. 

Although epidemic trees have been examined previously in specific data sets, their statistical properties have not been analyzed theoretically until now. We regard our results in that direction as among the main contributions of this paper. 

In a wider context, our approach suggests a possible starting point for a \emph{mesoscopic theory} of contagion, in which infection pathways, epidemic trees, and descendant distributions would play the leading role, operating at a scale in between the  local level of individual nodes and the global level of the entire network. 

To clarify these distinctions among the microscopic, mesoscopic, and macroscopic scales, consider the transition to a giant component in a susceptible-infected-removed (SIR) model of contagion on a network~\cite{newman2018networks, pastor2015epidemic}. Above the transition, there exists a giant infected component of size proportional to $N$.  
Such macroscopic phenomena have been extensively and fruitfully studied in the literature on network contagion \cite{newman2018networks, watts2002simple,lehmann2018complex,pastor2015epidemic}. But giant component sizes and other macroscopic quantities lump all infected nodes together, and thus discard information about which nodes infected which. Such causal information is retained in epidemic trees, which show the transmission pathways of contagion and thereby shed light on phenomena operating at the mesoscopic level. 

These mesoscopic considerations inescapably come into play (at least for mutant contagions on infinite-dimensional networks) because the descendant distribution is a beta function with a $d^{-2}$ tail, as we have shown above. A consequence of this inverse-square scaling is that the expected size of the \emph{mutant} infected component is of mesoscopic size comparable to $\log N$ for $N \gg 1$, and hence is intermediate in a precise sense; it is large compared to the $O(1)$ scale of individual nodes, but small compared to the $O(N)$ scale of the network itself, and of the giant infected (but non-mutated)  component. Note, however, that the variance of this smaller mutant infected component also diverges as $N \rightarrow \infty$. Hence its mean and variance do not adequately summarize the overall distribution of the number of mutant descendants, underscoring that one should rely only on the descendant distribution itself, as calculated here.  As a first step, the work presented here shows that descendant distributions are going to have an inverse-square tail on many real networks, even in the extreme limit where the SI model applies; away from this limit, if transmissibility is finite but still above the epidemic threshold, we expect descendant distributions to be this heavy tailed or even more so. One practical implication is that we should expect mutant strains of contagion to infect large fractions of network nodes occasionally. 

We expect that notions like contagion pathways, epidemic trees, and descendant distributions are just the beginning of a mesoscopic theory of contagion. Much remains to be discovered about the geometry and statistics of these and other quantities, both empirically for real contagions, and theoretically for a wide range of infection dynamics and network structures. Understanding this middle ground might also have practical benefits for the control of contagion processes, in contexts ranging from 
vaccination strategies for communicable diseases to methods for combating the spread of misinformation on social media. 

But before such practical benefits can be realized, any future mesocopic theory will also need to incorporate several realistic features that we have left out of the current model and analytical treatment. These include the extension to heterogeneities in degree, susceptibility, infectiousness, latency period, and so on. Such heterogeneities have shown themselves to be important in the COVID-19 outbreak~\cite{leclerc2020settings,miller2020full}, and are also thought to play a crucial role in the spread of many other infectious diseases \cite{galvani2005dimensions,lloyd2005superspreading}. Handling these heterogeneities theoretically will require extending the analytical treatment to a more sophisticated framework, like quenched mean-field theory~\cite{kiss2017mathematics, pastor2001epidemic}.

J.S.J. acknowledges funding through the University of Copenhagen UCPH 2016 Excellence Programme for Interdisciplinary Research, the Danish Council for Independent Research, and thanks the Center for Applied Mathematics at Cornell University for hospitality while this work was carried out. S.H.S. was supported by NSF Grant
CCF-1522054.
%
\clearpage
\renewcommand{\thefootnote}{\fnsymbol{footnote}}

\begin{center}
\begin{widetext}
  \textbf{\large Supplementary Information:\\[.1cm] Descendant distributions for the impact of mutant contagion on networks}\\[.3cm]
  Jonas S. Juul
  \\[.1cm]
  {\itshape \mbox{Niels Bohr Institute, University of Copenhagen, Blegdamsvej $17$, Copenhagen 2100-DK, Denmark}\\
  and \\
  Center for Applied Mathematics, Cornell University, Ithaca, New York 14853, USA}
  \\[.3cm]
  Steven H. Strogatz 
    {\\[.1cm]\itshape
  Center for Applied Mathematics, Cornell University, Ithaca, New York 14853, USA}
\\(\small Dated: \today)
\pagebreak
\end{widetext}
\end{center}
\setcounter{equation}{0}
\setcounter{figure}{0}
\setcounter{table}{0}
\setcounter{page}{1}
\renewcommand{\theequation}{S\arabic{equation}}
\renewcommand{\thefigure}{S\arabic{figure}}
\clearpage




\section{Complete graph}
In this section, we calculate the descendant distribution for a susceptible-infected (SI) process spreading on complete graph of $N$ nodes, with $N \gg 1$. 

Suppose that nodes are infected one at a time, and that the descendant distribution after $t$ nodes have been infected is given by $P_{d,t}$. We wish to calculate the equilibrium distribution of descendants, $P_d := \lim_{t\to\infty} P_{d,t}$. Note that when a new node is infected, a number of already-infected nodes will gain this node as a descendant. If, say, $14$ nodes acquire this node as a descendant, let us refer to this as introducing $14$ descendants in the epidemic tree and then distributing these $14$ descendants among the infected nodes. With this terminology in place, we proceed with the calculation.

First, because any edge that connects a susceptible and infected node is equally likely to be the next edge over which the infection is transmitted, and because the graph is complete, the expected \emph{fraction} of newly introduced descendants that nodes with $d$ descendants get is
\begin{equation}
    \frac{(d+1)P_{d,t}}{\sum_{d}(d+1)P_{d,t}} =  \frac{(d+1)P_{d,t}}{(m_t+1)} ,
    \label{eq:complete_firstEQ}
\end{equation}
where 
\begin{equation}
m_t := \sum_d dP_{d,t}
\end{equation}
is the mean number of descendants in the epidemic tree at time $t$. The numerator in Eq.~\eqref{eq:complete_firstEQ} expresses point (ii) in the main text, and the denominator is a normalisation factor. Next, to go from the expected fraction in Eq.~\eqref{eq:complete_firstEQ} to the expected \emph{number} of new descendants that a node with $d$ descendants gets in the following time step, we must multiply the expected fraction~\eqref{eq:complete_firstEQ} by the total expected number of new descendants, aggregated over nodes with \emph{any} number of descendants, that are added during the time step. 

To find this total, we observe that every infected node has equal probability of being the next to pass on the infection, and there are $t$ infected nodes at time $t$. Thus the probability that nodes with $d$ descendants will get a new descendant is $(d+1)P_{d,t}/t$. Summing over all $d$ then gives us the expected fraction of the infected nodes in total that will get a new descendant in the following time step; multiplying by $t$ gives us the corresponding expected number. This argument tells us, then, that
\begin{equation}
t\sum_d\frac{(d+1)P_{d,t}}{t} = m_t+1
\label{eq:total}
\end{equation}
is the expected number of infected nodes, in total, that will get a new descendant in the following time step. Note that the underlying network did not influence this last part of the calculation. 

By combining Eqs.~\eqref{eq:complete_firstEQ} and ~\eqref{eq:total} we find that, for the complete graph, the expected number of new descendants that a node with $d$ descendants gets in time step $t$ is 
\begin{equation}
\frac{(d+1)P_{d,t}}{(m_t+1)}(m_t+1) = (d+1)P_{d,t}.
\end{equation}
This result leads us to the following master equation, which expresses the expected gain and loss of nodes with $d$ descendants between time steps $t$ and $t+1$:
\begin{equation}
    (t+1)P_{d,t+1}-tP_{d,t} = 
    \begin{cases}
    1-P_{0,t} & \text{for } d=0,\\
    dP_{d-1,t}-(d+1)P_{d,t} & \text{for }d \ge 1. \\
    \end{cases}
\end{equation}
The case $d=0$ is different from other values of $d$ since the newly infected node will have no descendants when it is added to the epidemic tree, thereby making the gain term in the master equation equal to $1$. 
An equilibrium distribution must satisfy {$P_{d,t}=P_{d,t+1} =:P_d$}. Applying this condition and solving for $P_d$, we get:
\begin{equation}
    P_0 = \frac{1}{2}, \qquad
    P_{d} = \frac{d}{d+2}P_{d-1}.
\end{equation}
From this we conclude that the distribution of the expected number of descendants on the complete graph is
\begin{equation}
P_d = \frac{d!}{(d+2)!} = \frac{1}{(d+2)(d+1)}.
\label{eq:complete_graph_solutionSUP}
\end{equation}
As mentioned in the main text, keeping track of descendants can be mapped to growing a network by node copying. For the complete graph, this mapping means that Eq.~\eqref{eq:complete_graph_solutionSUP} is identical to the formula for the in-degree distribution calculated by Krapivsky and Redner~\cite{krapivsky2005network}. In their paper on network growth with node copying,  Krapivsky and Redner derive geometrical properties of the grown networks. We refer the interested reader to the paper, and continue with calculating descendant distributions for other classes of networks.


\section{Configuration model and {\ER} random networks}
\label{sec:configuration_model_and_ER}
Next we analyze two families of random networks: configuration-model networks, and {\ER}random graphs. 

In the configuration model that we consider, each of $N$ nodes has a certain number of ``half edges'' (or ``stubs'') sticking out of it, with the number of stubs being chosen at random from a prescribed degree distribution. The network is then generated by connecting pairs of stubs, chosen uniformly at random from the list of all stubs, to make the full edges of the resulting network. 

The \ER networks are constructed by  considering each pair of nodes independently and, with probability $p$, connecting that pair with an undirected edge. 

To calculate the descendant distribution for these random networks, we use the method of  the previous section. At an arbitrary time step $t\ge 1 $, an infected node with degree $k$ has at least one infected neighbor (its ``parent''). If the infected node (denoted \textit{I}), or one of its descendants, infects a neighbor on the next time step, then \textit{I} loses one edge over which it could infect another node. By doing this, however, it gets a new descendant, which might have a number of edges connecting it to susceptible nodes. If we assume that every one of the $k-1$ edges that could connect an infectious degree-$k$ node with a susceptible node has equal probability of doing so (equal to $1$ in the infinite-network limit), and if we assume that this probability is the same for every infected node, then an infected node has on average $(z-2)d+(z-1)$ edges which could connect it to susceptible nodes. Here $z$ is the mean degree of the network. 

So the mean number of new descendants that a node with $d$ descendants gets when a new node is infected is 
\begin{equation}
    \frac{\left[(z-2)d+z-1\right]P_{d,t}}{(z-2)m_t+z-1}\left(m_t+1 \right).
    \label{eq:ER_firsteq}
\end{equation}
Using this result, we can write down a master equation as we did when calculating the descendant distribution for the spreading process on the complete graph, and solve for a steady-state descendant distribution $P_d$, in the limit of infinite network size. After some algebra (see Section \ref{sec:universal_behavior} below for details, we find that 
\begin{equation}
    P_d = 
    \begin{cases}
    \frac{z-2}{2z-3} & \text{for } d= 0,\\
    \frac{z-2}{2z-3}\left[ B\left(\frac{z-1}{z-2},2 \right) \right]^{-1} B\left( \frac{z-1}{z-2}+d,2 \right) & \text{for } d\ge 1.
    \end{cases}
    \label{eq:ER_analyticalSUP}
\end{equation}
Here $B(a,b)$ is the beta function, which declines as $a^{-b}$ as $a\to \infty$ for fixed $b$. In our case, this means
\begin{equation}
\begin{split}
P_d &\propto B\left( \frac{z-1}{z-2}+d,2 \right), \\
&\propto d^{-2},
\end{split}
\end{equation}
for $d\gg 1$. By invoking identities for the beta function, we can rewrite the expression~\eqref{eq:ER_analyticalSUP} for the descendant distribution as
\begin{equation}
    P_d = \frac{z-1}{z-2}B\left( \frac{z-1}{z-2}+d,2 \right),
    \label{eq:Pd_conf}
\end{equation}
which is the expression we list in the main text. Figure~$3$ in the main text collapses the simulated data on the curve $B(\tilde{d},2)$, where 
\begin{equation}
    \tilde{d} := \tilde{x}(z) + d = \frac{z-1}{z-2}+d.
\end{equation}
Given a simulated data point $(d,P_d)$, this collapse is made by plotting the data point at $(d+\tilde{x}(z),\left[ \tilde{x}(z) \right]^{-1}P_d)$ instead.

\section{A ring and a small-world network}
\label{sec:small-world}
In both families of networks considered in the previous section, the edges are created according to a random procedure, and the resulting descendant distributions show the inverse-square scaling mentioned in the main text: $P_d \propto d^{-2}$ for large $d$. The question naturally arises whether this scaling law holds in complete generality, or whether it is restricted to certain  networks, and if so, what conditions imply it. 

It is easy to see that simple non-random graphs can display different limiting behavior for $P_d$. For example, consider a one-dimensional ring in which every node has only two neighbors, one to its left and one to its right. Then we can write down the descendant distribution immediately. Starting the contagion at a single seed, in each time step there will be exactly one possibility for the process to spread on the right hand side of the seed, and one possibility to spread on the left hand side. The resulting distribution of descendants in a ring consisting of $N$ nodes is 
\begin{equation}
P_d = \mathcal{N}\sum_{L=0}^N\big[\Theta\left(d \le L-1 \right)P_L+\Theta\left(d \le N- L+1 \right)P_L\big],
\end{equation}
where $\mathcal{N}$ is a normalization constant, $\Theta(x)$ is the Heaviside function equal to $1$ if $x\ge0$ and $0$ otherwise, and $P_L$ is the probability of the contagion process spreading exactly $L$ times to the left along the periphery of the ring, given by
\begin{equation}
P_L = \binom{N}{L}\left(\frac{1}{2}\right)^N.
\end{equation}

Another natural question to ask is then: How random does a network have to be to show the limiting behavior $P_d \propto d^{-2}$ seen earlier? In the rest of this section we analytically estimate the descendant distribution for the contagion process on small-world networks. Specifically, suppose the small-world networks are Newman-Watts small-world networks in which all nodes are connected to their two immediate neighbors on a ring lattice, and each node gets a shortcut to a neighbor chosen uniformly at random with probability $p$. 

First, we must estimate the expected number of new descendants that a node with $d$ descendants gets when a node gets infected. If the underlying network was simply a ring and no shortcuts had been added, every node would have equal chance of getting new descendants. This changes when the shortcuts are inserted: For each descendant a node has, the chance that one of its descendants has a shortcut increases. If the infection traverses such a shortcut link successfully, it can spread both to the right and to the left in this newly discovered part of the network. Hence, two more boundaries between infectious nodes and susceptible nodes have been created, and every node that has descendants on this boundary now has a higher chance of getting more descendants. This effect alters the expected number of descendants received by a node with $d$ descendants when a new node gets infected. The expected number now becomes 
\begin{equation}
\frac{P_{d,t}(1+2p(d+1))}{1+2p(m_t+1)}(m_t+1) .
\end{equation}
Here the first term represents the shortcut-independent probability that every node has to get a new descendant, and the terms that are proportional to $p$ correspond to the increased probability of getting new descendants that nodes get via shortcuts. 
With this, we can write down the master equation as for the complete graph and the random graphs of the section above. After some algebra (see Section \ref{sec:universal_behavior} below for details), we find
\begin{equation}
P_d =
\begin{cases}
\frac{2p}{1+4p} & \text{for }d=0, \\
\frac{2p}{1+4p}\left[B\left( \frac{2p+1}{2p},2 \right) \right]^{-1}B\left( \frac{2p+1}{2p}+d,2 \right) & \text{for }d\ge 0.
\end{cases}
\label{eq:SW_solutionSUP}
\end{equation}
For large $d$, this analytical solution declines as 
\begin{align}
P_d &\propto  B\left( \frac{2p+1}{2p}+d,2 \right), \\
&\propto d^{-2}.
\end{align}
In Fig.~1 of this Supplemental Material, we see that the analytical solution is indeed in qualitative agreement with the simulations.
\begin{figure}
    \centering
    \includegraphics[width=89mm]{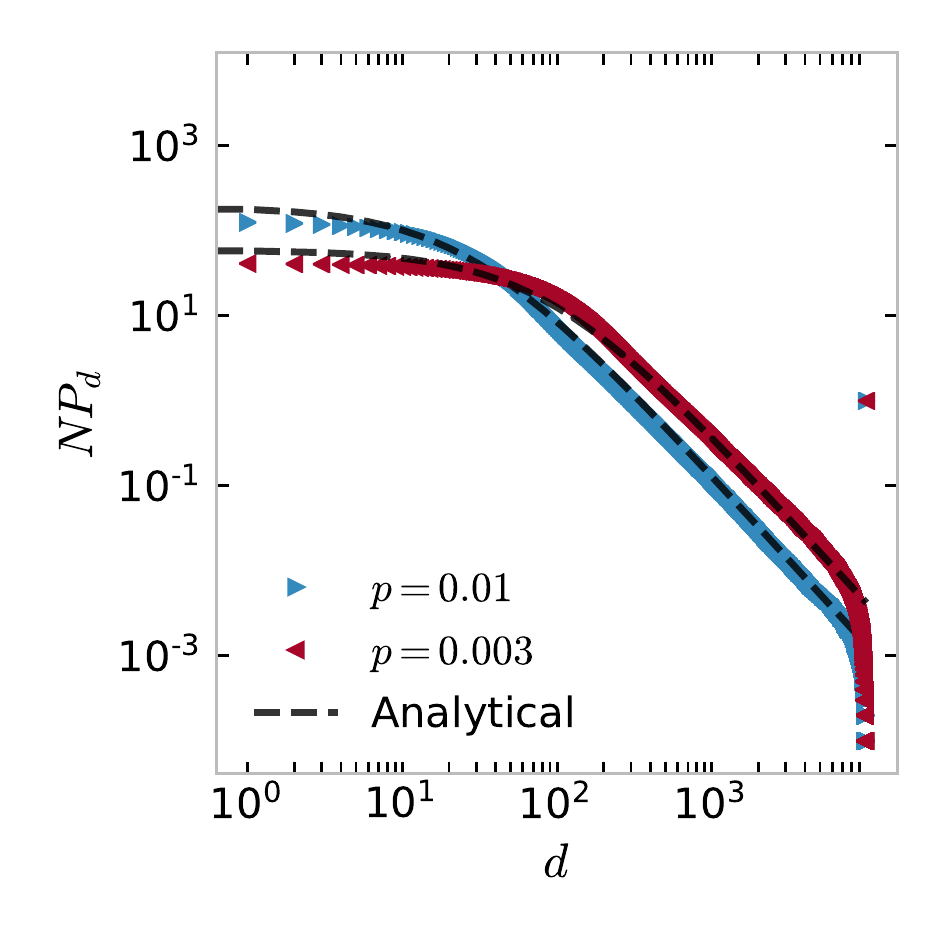}
    \caption{\textbf{Descendant distributions of an SI contagion process on Newman-Watts small-world networks.} The networks are created by starting out with a ring in which every node is connected to its two nearest neighbors, and then connecting each node with probability $p$ to another node chosen uniformly at random~\cite{newman1999renormalization}. The resulting descendant distributions, plotted here for networks of size $N=10^4$ nodes, show the same scaling behavior discussed in the main text: the distribution $P_d$ decays in proportion to $d^{-2}$ for large $d$, followed by a finite-size cutoff. We simulate the system for two values of $p$ and plot the resulting descendant distributions $P_d$ along with the analytical approximation~\eqref{eq:SW_solutionSUP} derived in Section \ref{sec:small-world}. }
    \label{fig:small_world}
\end{figure}
\section{Networks with block structure}
In the previous section, we provided analytical results for the descendant distribution arising from spreading in small-world networks. In the small-world network, the probability of a nodes having a shortcut, $p$, was crucial \--- with probability $p$, a newly infected node would open two new fronts of contagion transmission in another part of the network.

The same idea lets us calculate the descendant distribution arising from contagion in networks with a block-like structure. The network we consider consists of ``blocks''. In each block, nodes are connected to each other and have average degree $z$. The blocks could for example be \ER graphs or configuration-model graphs. In each of these $n$ blocks, every node has probability $p$ of receiving an additional edge leading to a node, chosen uniformly randomly, from the set of nodes located in other blocks.

We have already established that in a network with mean degree $z$, an infectious node with $d$ descendants has on average $(z-2)d+z-1$ edges which could connect itself or its descendants to susceptible nodes. This expression came from the fact that a node getting one additional descendant trades one edges potentially leading to susceptible nodes, while it gains $z-1$ new such edges. In the network with block-like structure, this is still true for spreading constrained to a single block.

If each node has probability $p$ of having an edge leading to a different block, infections traversing such an edge would provide, on average, $z$ edges to susceptible nodes in the new block. This is one more edge than would be gained by spreading internally in the block. Because a node with $d$ descendant could gain access to trans-block edges through itself or one of its descendants, such a node with $d$ descendants has on average $(z-2)d+z-1+p(d+1)$ edges which could connect itself or its descendants to susceptible nodes.

By repeating our calculating from Eq.~\eqref{eq:ER_firsteq}, we get the following expression for the descendant distribution in the network with blocks,
\begin{equation}
    P_d = \frac{z-1+p}{z-2+p}B\left( \frac{z-1+p}{z-2+p},2 \right).
\end{equation}
One again, the tail of the descendant distribution follows an inverse-square law.
\section{Scaling law and its breakdown}
\label{sec:universal_behavior}
We have studied the descendant distributions for simple contagion on various classes of networks: complete graphs, configuration-model networks, \ER networks, and small-world networks. In each case we proved that the descendant distributions decline as a power law with exponent $-2$ for large $d$. What structural feature accounts for this scaling law?

One thing that is true for all these graphs is that the probability of a node getting more descendants is linearly proportional to the number of descendants the node already has. In other words, the expected number of descendants received by a node with $d$ descendants, when a new node gets infected, is of the form
\begin{equation}
\frac{P_{d,t}\left( c+fd \right)}{\sum_d P_{d,t} \left(c+fd \right)}\left( m_t +1 \right) = \frac{P_{d,t}\left( c+fd \right)}{c+fm_t}\left( m_t +1 \right),
\label{eq:Pd_general_mt}
\end{equation}
for $c,f >0$. We will now show that this condition, along with $m_t\to\infty$ as $t\to\infty$, is sufficient to make the resulting distribution of the number of descendants decline as a power law with exponent $-2$ for large $d$. The condition $m_t \to \infty$ is true for all the classes of random graphs we have examined, since the decreased number of edges compared to the complete graph decreases the interface between susceptible and infectious nodes. This makes the probability of nodes with many descendants getting additional descendants increase compared to the spreading process on the complete graph. Because $m_t$ diverges for the complete graph, $m_t$ also diverges for the random graph in question by the comparison test.  As $m_t \to \infty$,  the right hand side of equation~\eqref{eq:Pd_general_mt} approaches $P_{d,t}(c/f+d)$. 

As in Section 1, this line of reasoning leads us to the master equation,
which expresses the expected gain and loss of nodes with $d$ descendants between time steps $t$ and $t+1$:
\begin{equation}
(t+1)P_{d,t+1}-tP_{d,t} = 
\begin{cases}
1-P_d \frac{c}{f} & \text{for }d=0,\\
\!\begin{aligned}
\bigg[ P_{d-1,t}\bigg(\frac{c}{f}+d-1\bigg)\\-P_{d,t}\bigg(\frac{c}{f}+d\bigg)\bigg]
\end{aligned}
& \text{for }d\ge 1.
\end{cases}
\end{equation}
Looking for steady-state solutions $P_{d,t+1}=P_{d,t}=:P_d$, we obtain
\begin{equation}
P_0 = \frac{f}{f+c}, \qquad P_d = \frac{c/f-1 +d}{1+c/f+d}P_{d-1},
\end{equation}
where the expression for $P_d$ is valid for $d\ge 1$. Denoting $c/f-1 =: \alpha$, we can use the recursive nature of the expression to rewrite $P_d$ as follows:
\begin{align}
P_d &= P_0\prod_{\lambda=1}^{d}\frac{\alpha+\lambda}{\alpha + 2+\lambda} \nonumber \\ 
&= P_0\frac{\Gamma(\alpha+3)\Gamma(\alpha+d+1)}{\Gamma(\alpha+1)\Gamma(\alpha+3+d)}.
\end{align}
If we increase the terms of the fraction by a factor of $\Gamma(2)$, and use the relation between gamma functions and beta functions, $\Gamma(x)\Gamma(y)/\Gamma(x+y) = B(x,y)$, we get
\begin{align}
P_d &= \frac{f}{f+c}\left[ B(\alpha+1,2) \right]^{-1}B(\alpha+d+1,2), \\
& = \frac{c}{f}B\left(\frac{c}{f}+d ,2\right).
\end{align}
The final step was made by inserting the value of $\alpha$ and evaluating $B(c/f,2)=f^2/[c(c+f)]$. The asymptotic behavior for large $d$ is
\begin{align}
P_d &\propto (c/f+d)^{-2} \nonumber\\
&\propto d^{-2}.
\end{align}
Therefore, if the probability of getting more descendants increases linearly with the number of descendants a node already has, the descendant distribution will decline as $d^{-2}$ for large $d$. If we interpret the number of descendants of an infected node as a volume, and the interface separating infectious and susceptible nodes as a surface area, the descendant distribution will show the observed scaling if the surface area and the volume increase equally fast (proportional to $d$); in other words, if the network is infinite dimensional.

On this basis, we also expect that the $d^{-2}$ scaling should break down for networks  whose dimensionality remains  finite as the number of nodes tends to infinity. Such networks include one-dimensional chains, two-dimensional grids, and three-dimensional lattices. Indeed, simulations of the model contagion on two-dimensional square grids support this prediction: descendant distributions deviate significantly from the $d^{-2}$ scaling, as shown in Fig.~2 in this Supplemental Material. 
\begin{figure}
    \centering
    \includegraphics[width=89mm]{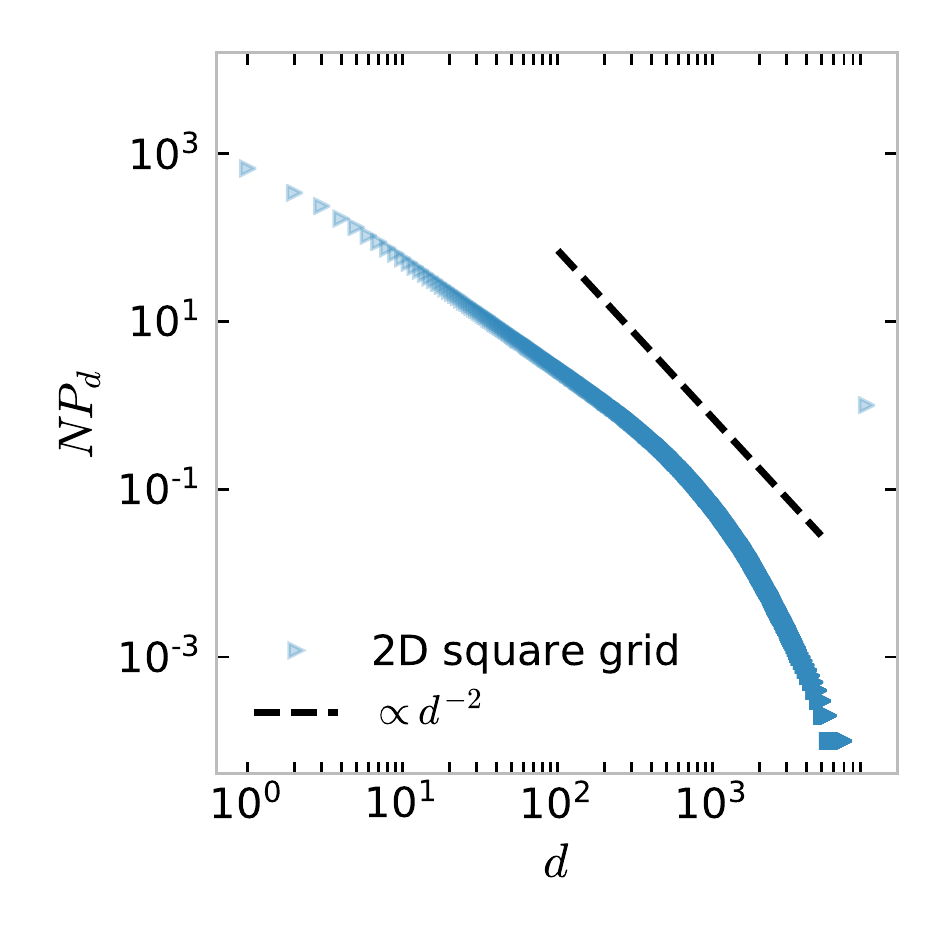}
    \caption{\textbf{Descendant distributions of an SI contagion process on two-dimensional square grids with periodic boundary conditions.} The networks consist of $N=99^2$ nodes, and $10^3$ random realizations of the spreading process were simulated. Because the underlying network is two-dimensional rather than infinite-dimensional, the resulting descendant distributions do not show the scaling law  discussed in the main text: the distribution $P_d$ does not decay in proportion to $d^{-2}$ for large $d$.}
    \label{fig:2Dsquaregrid}
\end{figure}
\section{Scale-free networks}
\begin{figure}
    \centering
    \includegraphics{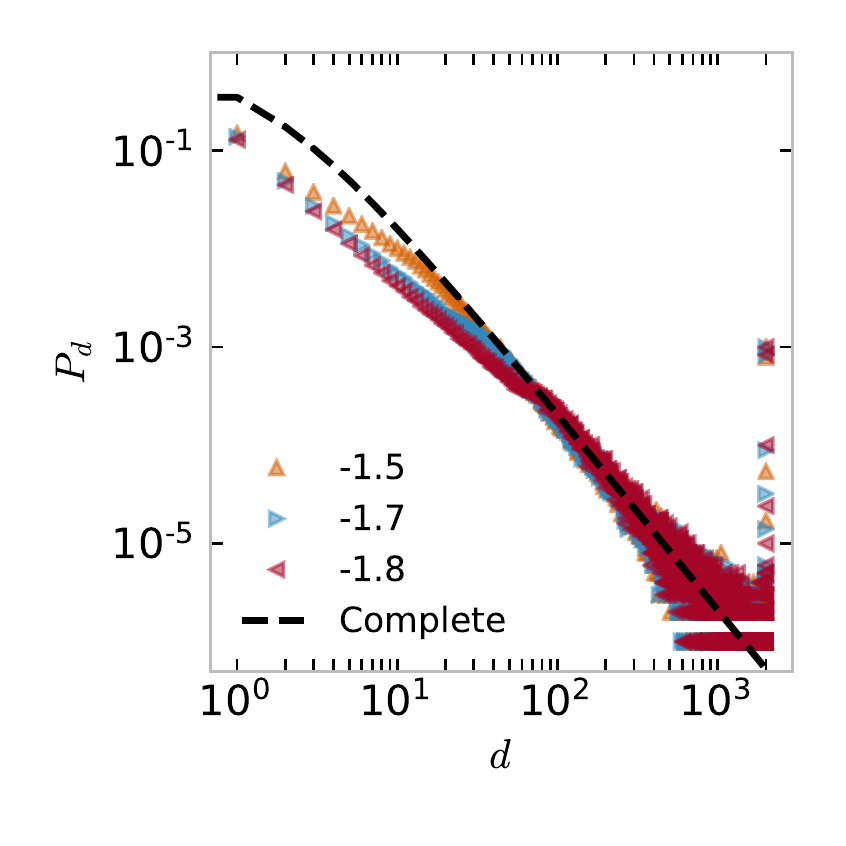}
    \caption{\textbf{Descendant distributions for an SI process on scale-free networks with different exponents in the degree distribution}. We constructed  scale-free networks with degrees drawn from power-law probability distributions with three different exponents (see figure legend).  We started a contagion at a random seed node, and let exactly $2000$ nodes get infected. Then we stopped the spreading, obtained the descendant distribution for the realization, and started a new simulation with a seed chosen uniformly at random. The decline in the distribution occurs slowly at first, and then more rapidly, revealing an apparent crossover. The resulting descendant distributions take longer to cross over to the inverse-square decline if the degree exponent is more negative.}
    \label{fig:ScaleFree}
\end{figure}

Networks with heavy-tailed degree distributions have received much attention~\cite{barabasi1999emergence,clauset2009power,broido2019scale-free,gerlach2019testing}. These heavy-tailed degree distributions can give rise to interesting dynamics in relation to contagion, for example a disappearing epidemic threshold ~\cite{pastor2001epidemic}. We simulate $10^4$ instances of the simple SI contagion process spreading on three different configuration-model scale-free networks. We drew between $11,000$ and $13,000$ degree values from probability distributions declining proportional to $k^{-\alpha}$ (with cutoff at $k=3\cdot 10^4$). We then selected a seed uniformly at random from the nodes located in the largest connected component, let $2,000$ nodes become infected, recorded the descendant distribution, and repeated this process $10^3$ times. For the values $\alpha \in \{-1.5,-1.7,-1.8 \}$ the largest connected component consisted of no less than $10^4$ nodes. Figure~\ref{fig:ScaleFree} compares the resulting descendant distributions to the analytical solution obtained for the complete graph.

Interestingly, the descendant distributions look qualitatively different from those studied in the main text, where all the networks lacked hub-like structure. Here, for the three scale-free networks, all three distributions seem to decline slowly for small values of $d$ before eventually declining more rapidly. Furthermore, as the degree exponent $\alpha$ becomes more negative, it takes longer for the distribution to converge to the faster decline, and the transition seems to happen close to where the simulated distributions intersect with the analytical solution for the complete graph. Understanding this seeming crossover behavior in detail would be an interesting future direction for research. How exactly does hub structure decrease the effective dimension of networks? 

\bibliography{bibliography_descendants}
\end{document}